\documentclass[12pt]{article}
\usepackage{amsmath,graphicx}

\def\EEE{E_1{}^1}

\def\be{\begin{equation}}
\def\ee{\end{equation}}

\def\s{{\rm sech}(w\tau)}
\def\t{{\rm tanh}(w\tau)}

\def\Sm{\Sigma_-}
\def\Sc{\Sigma_\times}
\def\Nm{N_-}
\def\Nc{N_\times}
\def\Sp{\Sigma_+}

\begin{document}

\begin{center}
{\Large\bf Generating matter inhomogeneities in general relativity}
\vspace{.3in} \\ 
{\bf A A Coley}, 
\\Department of Mathematics \& Statistics, Dalhousie University,\\
Halifax, Nova Scotia, Canada B3H 3J5
\\Email: aac@mathstat.dal.ca
\vspace{.1in}
\\ {\bf W C Lim},
\\Albert-Einstein-Institut, Am M{\"u}hlenberg 1, D-14476 Potsdam, Germany
\\Email: wclim@waikato.ac.nz
\vspace{.1in}
\vspace{0.2in}

%%\today
\end{center}

%%[PACS: 98.80.Jk,04.50.+h]

\begin{abstract}

In this Letter we discuss a natural general relativistic 
mechanism that causes inhomogeneities and hence generates matter
perturbations in the early universe.  We concentrate on spikes, both
incomplete spikes and recurring spikes, that naturally occur in the
initial oscillatory regime of general cosmological models.  In
particular, we explicitly show that spikes occurring in a class of
$G_2$ models lead to inhomogeneities that, due to
gravitational instability, leave small residual imprints on matter in the
form of matter perturbations.  The residual matter overdensities from
recurring spikes are not local but form on surfaces.  We discuss the
potential physical consequences of the residual matter imprints 
and their possible effect on the subsequent
formation of large scale structure.

\end{abstract}

%%\newpage

{\bf{Introduction}}.
It is a general feature of solutions of partial differential 
equations (PDE) that spikes {\footnote{Spikes are narrow inhomogeneous 
structures (see, e.g., Figure 2). }} occur
\cite{spikes}.  Therefore, spikes are expected to occur in generic
solutions of the Einstein field eqns (EFE) of general relativity (GR),
which are a complicated set of PDE \cite{thesis}.  
Indeed, when a solution of the EFE is stable at certain points but otherwise
unstable, spikes will arise near the stable points.  
The set of such
points can be a surface, a curve or a point in the three-dimensional
space.

Belinskii, Khalatnikov and Lifshitz
(BKL) \cite{BKL} have conjectured that within GR, the approach to the
generic (past) spacelike singularity is vacuum dominated, local, 
and oscillatory (i.e., Mixmaster).
Studies of $G_2$ 
{\footnote{The so-called $G_2$ spacetimes are those
which admit two commuting spacelike Killing vector fields, and hence only have one degree
of inhomogeneity.}}
and more general cosmological models have 
produced numerical evidence that the BKL conjecture generally 
holds except possibly at isolated points 
(surfaces in the three-dimensional
space) where spiky
structures (`spikes') form \cite{G2}. These spikes become ever
narrower as the singularity is approached. 
The presence of such spikes violates the local part of the BKL conjecture.

The study of spikes is
severely limited due to the enormous numerical resources
needed to resolve the narrowing spikes in simulations. In 
\cite{G2num}, further improved numerical
evidence was presented that spikes in the Mixmaster 
regime of $G_2$ cosmologies are transient and recurring, 
supporting the conjecture that the generalized Mixmaster 
behavior is asymptotically non-local where spikes occur.
It is believed that this recurring violation of BKL locality holds in
more general spacetimes;
however, it remains to study this more comprehensively.

In this Letter, we wish to study the residual imprints of the spikes
on matter inhomogeneities. As the spike inhomogeneities form,
matter undergoes gravitational instability and begins to collapse to form overdensities. In this way, density inhomogeneities
naturally form in the early universe in generic GR models.
This is a natural GR effect, and the question arises as to whether these
density inhomogeneities are physical and can be observed and, indeed, whether such matter inhomogeneities could act as seeds for the subsequent formation of large scale structure.
{\footnote{ We shall use the general terminology matter inhomogeneities or 
density perturbations
or overdensities to describe the residual local inhomogenieties imprinted 
in the matter, but the actual mathematical behaviour is rather 
more complicated and perhaps can be better descibed as a delay 
in evolution at the spike (also note that $\Omega$ and the physical
fluid density $\rho$ may have 
different qualitative behaviour, due to the evolution of the Hubble
normalization).}}

Therefore, we are interested in possible (purely classical GR) effects that could
cause inhomogeneities and hence generate matter perturbations that might then impinge
on structure formation.  We are particularly interested in recurring and distributed
spikes formed in the oscillatory regime (or recurring spikes for short), and their
imprint on matter.  We show that matter inhomogeneities can occur in simple so-called
orthogonally transitive (OT) $G_2$ models \cite{G2} with a tilted radiation fluid.
{\footnote{A tilted fluid is a fluid that has a non-zero velocity (or tilt) relative
to the chosen reference frame.}} Some intuition can be gained by looking at the case
of small (normalized) density parameter $\Omega$ with small tilt.  First, we study a test fluid in an exact vacuum
background \cite{exactspike} (small tilt and small $\Omega$ with no backreaction on
the geometry); although a complete spike transition leaves no imprint in this case, a
partial spike transition does leave an imprint.  We then study the small tilt
approximation with small $\Omega$ (when matter is not a test field), and show
heuristically that residual matter inhomogeneities occur.  The argument doesn't apply
to the typical case of non-negligible $\Omega$ and/or tilt; however, we argue it does
occur in more general models (partly by studying the decay rates of the inhomogeneous
spikes).  We also discuss some simple exact spike solutions (previously studied in
the literature), which have residual matter inhomogeneities, further lending support
to the heuristic analysis \cite{thesis}.  But ultimately we shall demonstrate the
existence of matter perturbations by numerical simulations.

Both the incomplete spikes and the recurring spikes are potentially of physical importance 
(but perhaps for different reasons). 
We are particularly interested in the ($G_2$) recurring spikes, and we
show explicitly that there exist spikes leading to
inhomogeneities and a small residual in the form of matter perturbations. Moreover,  these 
density perturbations
occur naturally within generic cosmology models within GR and they are not
local (or point-like) but form on surfaces and give rise to a distribution of perturbations.
These classical GR generated inhomogeneities consequently may affect large scale structure.
Larger effects 
(e.g., occuring from incomplete spikes), must of course be consistent with observations.
 
In the final section we discuss the potential physical applications of the results, and
speculate on whether these 
recurring spikes might be an alternative to the inflationary mechanism for 
generating matter perturbations.

{\bf{Spike Analysis}}.
The full evolution (EFE) equations are given in Appendix D of \cite{thesis} 
(with $A=0$).
The variables there are $|\beta|$-normalized (or Hubble--normalized). The 
$\beta$-normalized lapse is chosen to have  
the value $-1/2$ here
(which is appropriate for numerical simulation).
The evolution  equations are:
{\footnote{$\beta$ is the area expansion
rate of the $(y,z)$ plane; $\gamma$ is the equation of state parameter,
with $\gamma=\frac13$ describing the radiation fluid.  $\Sp$, $\Sm$
and $\Sc$ are components of the $\beta$-normalized rate of shear;
$\Nc$ and $\Nm$ are components of the $\beta$-normalized spatial
curvature.  $\Omega$ is the $\beta$-normalized fluid density; $v$ is the
relative fluid velocity (tilt) in the $x$-direction.
$\beta$-normalization is analogous to the standard
Hubble-normalization, and is related through $H$ = $\beta(1-\Sp)$.}}
\begin{eqnarray}
\partial_\tau \ln|\beta| &=  &\frac34 [1+\Sm^2+\Sc^2+\Nc^2+\Nm^2 + (\gamma -1) \Omega]\\
\partial_\tau \ln\Omega & = & \frac12 \gamma v  \EEE \partial_x \ln\Omega + \frac12
\gamma \EEE \partial_x v \nonumber\\
&& - \frac34 (2 - \gamma) [1 + \Sm^2 + \Sc^2 + \Nc^2 + \Nm^2 - \Omega],
\end{eqnarray}
where $\Sp$ and $q$ (and all terms) are given in \cite{thesis}.
There are also evolution equations for $\EEE$ and $v$ (and $\Sm, \Sc, \Nm, \Nc$).
{\em{In this Letter we need only focus on $\Omega$ and $v$.}}
We consider the linearization of each variable around a background metric: 
\be
	\Omega = \Omega_0 + \epsilon \Omega_1 + O(\epsilon^2),\quad
	v = v_0 + \epsilon v_1 + O(\epsilon^2)
\label{other_expansion}
\ee
(etc.). In the small $\Omega$ and small $v$ approximation we assume that $\Omega$ 
and $v$ have 
vanishing zeroth order terms.
The linearized evolution equations are easily obtained from the full EFE.

The expressions for the exact  vacuum spike solution \cite{exactspike}, which are used as the
zeroth order (background) solution in the linearization are:
\be
\label{spike}
        (\Sm,\Nc,\Sc,\Nm) =
        \left(-c \Sm{}_\text{Taub} -\frac{1}{\sqrt{3}},
        s\Nm{}_\text{Taub},
        c \Nm{}_\text{Taub},
        -s \Sm{}_\text{Taub}
        \right).
\ee
\be
\label{csf}
        c = \frac{f^2-1}{f^2+1},\quad
        s = \frac{2f}{f^2+1},\quad
        f =w e^\tau \s x.
\ee
\be
        \Sm{}_\text{Taub}=\frac{w}{\sqrt{3}}\t-\frac{1}{\sqrt{3}},\quad
	\Nm{}_\text{Taub}=\frac{w}{\sqrt{3}}\s
\ee
More importantly, we note that
\be
        \beta = -\frac12 \s e^{\frac{w^2+7}{4}\tau - \frac14 \lambda_2} (f^2+1)^{-\frac12}.
\label{beta}
\ee

We shall provide analytical evidence for the existence of residual matter inhomogeneities
due to spikes. We study
the case of negligible Hubble-normalized density $\Omega$ and negligible tilt $v$.
To see the effect of a spike on matter, we first consider a test perfect fluid.

{\em{Spike imprint: test fluid and incomplete spikes}}:
For a test fluid (with negligible $\Omega$ and $v$), from
the EFE  and the conservation equation we obtain:
\begin{align}
	\partial_\tau \ln|\beta| &= \frac12 (q+1)
\\	
	\partial_\tau \ln\Omega &= - (q+1) + \frac32\gamma (1-\Sp).
\end{align}
Remarkably, ($\rho$ and) $\Omega$ can be solved exactly as follows:
\be
\label{imprint}
	\rho = \rho_0(x) \left( \frac{\beta}{\beta_0(x)} \right)^\gamma,
\quad   
	\Omega = \Omega_0(x) \left( \frac{\beta}{\beta_0(x)} \right)^{-(2-\gamma)}.
\ee

Note that $\partial_\tau \ln|\beta|$ is positive.
So for $\gamma$ satisfying $0<\gamma<2$, 
we see that $\beta$ and $\rho$ blow up at the singularity, 
but $\Omega = \rho/(3\beta^2)$ tends to zero [``matter does not matter"].
As spacetime expands, $\rho$ and $\beta$ decrease and $\Omega$ increases.

For the spike solution, $\beta$ is given by
(\ref{beta}), where
\be
\label{eqn11}
	(f^2+1)^{-\frac12} \equiv [(w e^\tau \s x)^2 + 1]^{-\frac12},
\ee
depends on $x$ and so it is responsible for the spatial inhomogeneity.
It can be seen from Fig. 1 that this factor is asymptotically homogeneous 
(and equals $1$) 
as $\tau$ tends to $\pm \infty$, for $|w|>1$.
It takes the value $1$ at $x=0$ for all time, but is smaller for $x\neq0$.
This means that $\beta$ is inhomogeneous during a spike transition,
but a complete spike transition restores $\beta$ to homogeneous.
So the cumulative effect of a complete spike transition on spatial inhomogeneity in $\beta$ is zero.
Similarly, a complete spike transition has zero cumulative effect on the 
inhomogeneity in $\rho$ and $\Omega$ (when regarded as a test fluid).

\begin{figure}[t!]
  \begin{center}
    \resizebox{11cm}{!}{\includegraphics{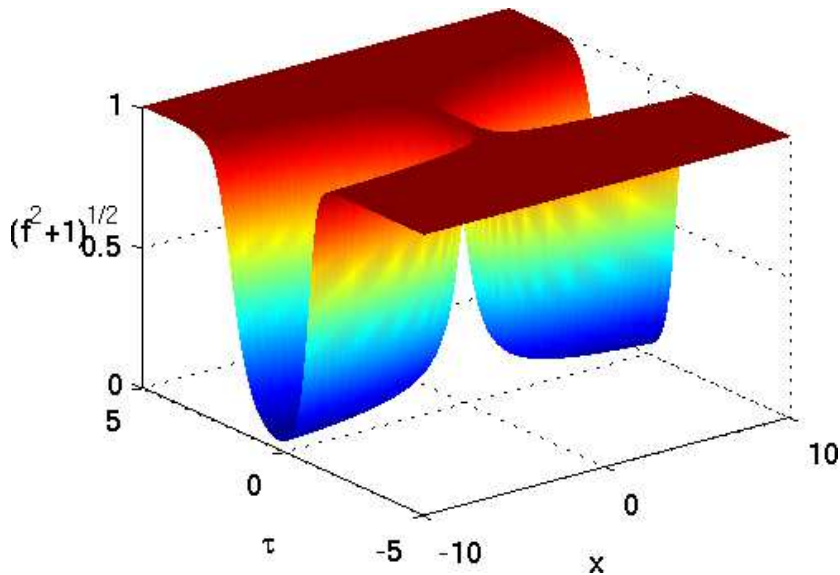}}
    \caption{The factor $(f^2+1)^{-\frac12}$ with $w=3$.
    $x$ and $\tau$ are the dimensionless space and time variables,
respectively.  $\tau$ tends to $+\infty$ at the singularity.  The
spatial axis used here (and in Figures 2 and 3) is $X=e^\tau x$.}
    \label{fig_fac}
\end{center}
\end{figure}

%%%%\resizebox{\linewidth}{!}{\includegraphics{fac.eps}}

{\em{Discussion}}:
It is interesting that for a partial spike transition (i.e., during a spike transition), 
$\beta$ develops a hump at the spike surfaces 
$x=0$. Correspondingly, $\rho$ develops a hump at $x=0$, and $\Omega$ develops a dip 
at $x=0$.
This can be interpreted as a delay or lag in the evolution at the spike point compared with points far away. 
The delay is temporary, and by the end of the spike transition, there is no delay.

\begin{figure}[t!]
  \begin{center}
    \resizebox{7cm}{!}{\includegraphics{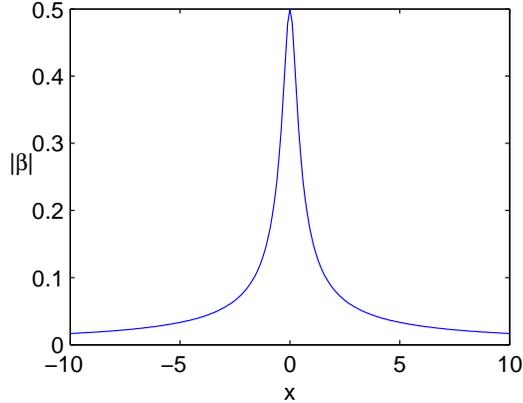}}
    \caption{$|\beta|$ at $\tau=0$ with $w=3$, $\lambda_2=0$.}
    \label{fig_absbeta}
\end{center}
\end{figure}

As the universe expands from the big bang, presumably it starts with a Mixmaster oscillatory regime
with spike transitions occuring on surfaces here and there, 
and with Bianchi type II transitions occuring elsewhere.
Eventually, the oscillatory regime ends when $\Omega$ is no longer negligible.
Some of the spike transitions are in the middle of transitioning when the oscillatory regime ends.
This leaves an inhomogeneous imprint on matter as well as curvature, by way of a delay in evolution at surfaces of spike points.
$|\beta|$ and $\rho$ are larger, and $\Omega$ is smaller at surfaces of spike points than at other points.

Since  $\Omega$ is no longer negligible at the formation of an incomplete spike, eqn.
(\ref{imprint}) may not be a good approximation.
But the conclusion that $|\beta|$ and $\rho$ are larger, 
and $\Omega$ is smaller at surfaces of spike points, is confirmed
by numerical simulations.
We note that the isolated incomplete spikes are rare, since the solutions spend much less time undergoing transitions than 
evolving in the Kasner epoch.

{\em{Heuristic analysis: small $\Omega$}}.
We next study the linearized equations with a spike background, 
assuming a small $\Omega$.
The zeroth order terms in the linearized equations are satisfied 
identically by the exact spike solution.
Assuming $\Omega_0 =0$, then
the leading order term in the linearized EFE gives $\Omega_1$ in terms of $\beta_0$
(eqn. (\ref{beta})):
\be
        \Omega_1 = \hat{\Omega}_1(x) \left( \frac{\beta_0}{\hat{\beta}_0(x)} \right)^{-(2-\gamma)},
\ee
where $\hat{\Omega}_1(x)$ and $\hat{\beta}_0(x)$ are the value at a fixed time. Without loss of generality, we can 
evaluate $\hat{\Omega}_1(x)$ and $\hat{\beta}_0(x)$ at $\tau=0$. We then obtain
\be
        \hat{\Omega}_1 = C (w^2 x^2 + 1)^{-\frac{\gamma(2-\gamma)}{2(\gamma-1)}}  |\hat{\beta}_0|^{-(2-\gamma)},
\ee
where $C$ is a constant.
This means the initial spatial profile of $\Omega_1$ is not freely specifiable (for non-dust), but is determined by the spike background.

We can then solve for $(\EEE)_1$.
The other unknowns, namely $\beta_1$, $(\Sm)_1$, $(\Sc)_1$, $(\Nm)_1$ and $(\Nc)_1$, are coupled, although
$v_1$ decouples from these unknowns. However, even in this case, where we assume that
both $\Omega$ and $v$ are small,
we must ultimately resort to numerical analysis.

{\em{Heuristic analysis: non-trivial $\Omega$}}.
Let us now consider the large $\Omega$ case with $\Omega_0 \neq 0$.
From these eqns (for $v_0$ small) we obtain
$$ \partial_\tau \ln \Omega = \partial_\tau \ln |\beta|^{-2-\gamma} + 
\frac{3}{4}\gamma (2 - \gamma) \Omega. $$
Writing $\Omega = \Omega_0 (1 + \epsilon \Omega_1)$, where
$\Omega_0 \equiv \hat{\Omega}_0(x)  \beta_0^{-(2-\gamma)}$ is the zeroth order solution as given by
eqns. (\ref{imprint}) and (\ref{beta}), we obtain
$$\partial_\tau (1 + \epsilon \Omega_1) = \frac{3}{4} \gamma (2 - \gamma) \Omega_0
(1+\epsilon \Omega_1)^2, $$
and hence
$$ \Omega_1 = \hat{\Omega}_1(x) - \frac{4}{3 \gamma(2-\gamma)} \bigg[ \int \Omega_0 d \tau  \bigg]^{-1}  $$
where $\Omega_0 = \hat{\Omega}_0 (x) F(\tau) B(\tau, x)$, and $B(\tau, x) \equiv 
(1+f^2)^{1-\gamma/2}$
(and $F(\tau)$ is defined by eqn. (\ref{beta}) and $f$ is defined by (\ref{eqn11})),
plus possible additional contributions from $\beta_1$.
The important point is that there are consequently contributions to $\Omega_1$ from terms like
$\int  B (\tau, x)d \tau$.  Now, the spike occurs for $-\tau_0 < \tau < \tau_0$, but as
$\tau \to \infty$ the spike disappears and $B(\tau, x)$ becomes homogeneous
(although transient inhomogeneities occur for $-\tau_0 < \tau < \tau_0$, see Fig. 1)
However, terms like $\int B(\tau, x) d\tau$ retain a residual inhomogeneity.  
Hence terms like $\int B(\tau, x)
d\tau$ (and integrals thereof) contribute an inhomogeneous imprint to $\Omega$, and hence the
matter density.  This is illustrated  in Fig. 3  (for the same B as in Fig. 1).

Therefore, there will be a residual inhomogeneous imprint on the density due to a spike.  This has been illustrated
here in this simple case, but it is most likely to be true in all generality.  
We still need to analyse the case of large $\Omega$ and large $v$. However, this
can only be done numerically.

\begin{figure}[t!]
  \begin{center}
   \resizebox{7cm}{!}{\includegraphics{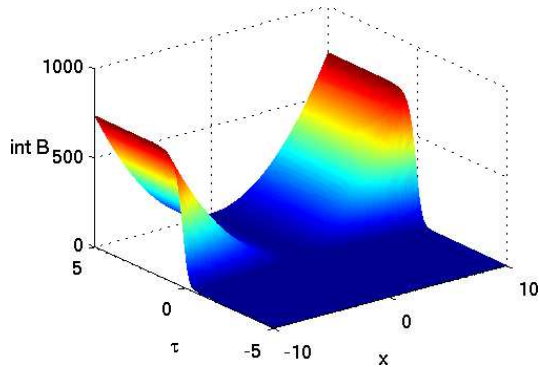}}
    \caption{Integral of B; see Fig. 1.}
    \label{intB.eps}
\end{center}
\end{figure}

%\begin{figure}[t!]\begin{center}\resizebox{\linewidth}{!}{\includegraphics{intintB.eps}}\caption{Integral of intB; see Fig. 1.}\label{intintB.eps}\end{center}\end{figure}

%\begin{align}
%	\Sp &= \frac12(1-\Sm^2-\Sc^2-\Nm^2-\Nc^2 \Omega)
%\\
%	q &= \frac12 + \frac32(\Sm^2+\Sc^2+\Nm^2+\Nc^2} + \frac32(\gamma-1)\Omega
%\end{align}

%\begin{align}
%\partial_\tau (\ln\EEE) = -1 + \frac34(2-\gamma)\Omega\\
%\partial_\tau \ln|\beta| &= \frac34(1+\Sm^2+\Sc^2+\Nm^2+\Nc^2 + 2(\gamma-1)\Omega)\\
%\partial_\tau \ln\Omega &= STUFF + \frac34(2-\gamma)
% (1+\Sm^2+\Sc^2+\Nm^2+\Nc^2 -\Omega)\\
%\partial_\tau \ln\v &=	STUFF
%\end{align}]]

{\em{Decay rate}}:
From above $\Omega_0 = \hat{\Omega}_0 (x) F(\tau) B (\tau,x)$, where 
 $$B(\tau, x) = [1+ (w e^\tau {\s}   x)^2]^{1-\gamma/2}  $$
$(\gamma <2)$.  The residual inhomogeneity comes from $B(\tau, x)$.
Of course, as $\tau \to \infty$, $B(\tau, x) \to 1$, and the inhomogeneity
from the spike decays.  For large $\tau$ (or equivalently for $\tau \to -\infty$), we have
that
$$ B(\tau, x) \sim 1 + 4w^2 (1 - \frac\gamma 2) e^{2(1-w)\tau} x^2$$
where $w > 1$ (assuming $w$ positive).  Hence the inhomogeneity 
(the imprint of the spike) decays at the rate $e^{2(1 - w)\tau}$.

Now, from linear perturbation theory, density inhomogeneties are expected to grow at the rate $t^c \sim
e^{c \tau}$, for some constant $c$, due 
to gravitational instability (where $c$ depends on $\gamma$;
for example, and simply as a reference, in a flat Friedmann-Lemaitre (FL) model, $c= (\gamma - \frac 1 3)$ on the largest scales).  Thus, it might be 
expected that residual
inhomogeneous imprints on the density would occur whenever $c + 2(1 - w) >0$
(i.e., for $1 < w < 1 +c/2$).

{\em{Exact solutions with matter; LRS $G_2$ dust and LTB models}}:
Finally, we briefly discuss some simple exact spike solutions 
that have been studied previously,
which constitute examples of residuals in special cases.

Spikes arise when a solution straddles the
stable manifold (or separatrix) of an unstable equilibrium point (a
source or a saddle point) \cite[Chapter 6]{thesis}.
If the unstable manifold of that equilibrium
point is one-dimensional, then spikes occur on a surface (in
three-dimensional space); if two-dimensional, then on a curve; if
three-dimensional, then on a point.

In \cite{thesis}, an explicit solution in the class of locally-rotationally 
symmetric (LRS)
$G_2$ cosmological models with dust and a cosmological constant (a special
case of the Szekeres solution) was given,
illustrating explicitly how the straddling of the stable manifold of a saddle point leads to the
formation of a spike.  Solutions on each side of the stable manifold approach different
sinks, leading to a discontinuous limit in some of the variables.  The shape of this
discontinuous limit may look like a step function or a spike. The explicit form for the
inhomogeneous $\Omega(x)$ (with $\Lambda = 0$), which develops a step/jump inhomogeneity,
was given in \cite{thesis}.
Similarly, it is possible to find a spiky solution in Lemaitre-Tolman-Bondi (LTB) models 
\cite{thesis}, which may be more relevant 
to structure formations than LRS $G_2$ models (we shall persue this further in future work).

%%\newpage

{\bf{Numerical evidence of residuals}}.
Let us do a simple numerical simulation to show that there is indeed an 
inhomogeneous imprint left by a spike.
The numerical code is essentially the one used in ~\cite{G2num}, with a
radiation fluid added.
The limitation of the code is that it cannot handle shock waves and the
step-like structure that forms in the tilt.
The tilt is unstable for $\Sigma_+ < 0$, so this limits the simulations
to spikes with roughly $|w| < 2$.
The current code with zooming also means that each simulation can only
see the spacetime at horizon scale.
To see the imprint, which becomes super-horizon into the past, several
simulations are needed to produce the figure.
It is inefficient, and we hope to find an efficient way to simulate the
spacetime at super-horizon scale in the future.

\begin{figure}[t!]
  \begin{center}
    \resizebox{12cm}{!}{\includegraphics{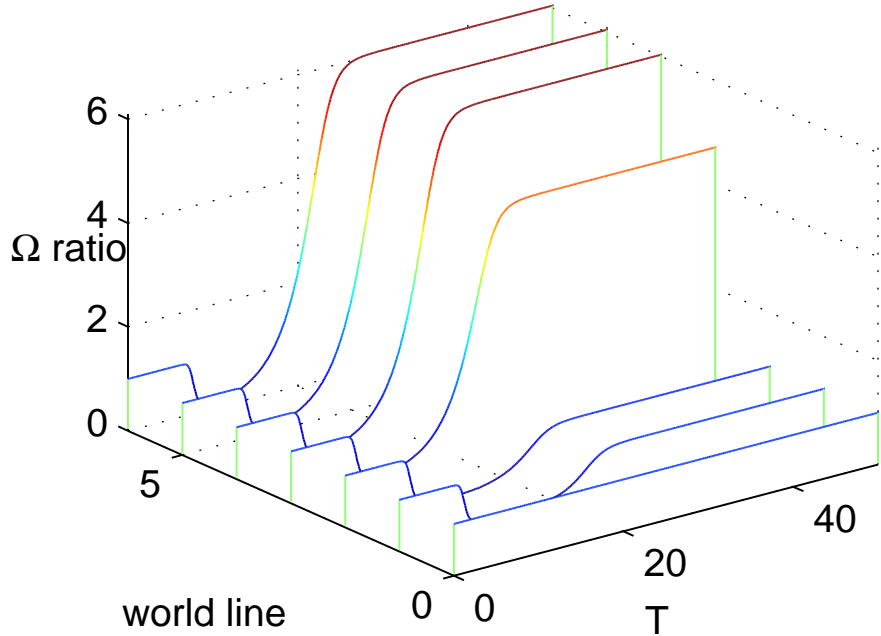}}
    \caption{Ratio of $\Omega$ along different worldlines. See text for details.}
    \label{fig_waterfall}
\end{center}
\end{figure}

Fig.~\ref{fig_waterfall} shows the time evolution of 7 separate simulations centred on
7 worldines
\be
        x = 0,\ 10^{-7},\ 10^{-6},\ 10^{-5},\ 10^{-4},\ 10^{-3}\ \text{and}\ 10^{-2}
\ee
(labeled worldline number 6 to 0 in the plot). The ratio of $\Omega$ is taken to be
\be
        {\Omega}/{\Omega|_{x=10^{-2}}}.
\ee
Hence the plot along woldline number 0 equals 1.  From the plot, towards the singularity we
see that $\Omega$ near the spike (at $x=0$) grows 6-fold.  Reversing the time direction,
we see that the spike thus leaves an imprint on $\Omega$ with a 6-fold underdensity.  The above
plot is produced with $\Omega=10^{-10}$ and $w=1.5$.  Using $w=1.4$ yields 
a $5.8$-fold change.  Using
$\Omega=10^{-5}$ does not affect the result.  From these preliminary results, we conclude
that the spike leaves an underdense imprint on $\Omega$ regardless of the size of $\Omega$.
The amplitude of the underdensity depends on $w$.

Let us make  two comments about the imprint obtained. First, 
the wavelength is at the width of the spike (roughly the horizon scale) when
it is created. But towards the future it will become sub-horizon, as the horizon
expands.  Because of this, it is difficult to see
the imprint numerically when using a single simulation with zooming. Second,
the imprint is large (6-folds) because $\Omega$ is close to zero. Towards the
future this ratio will become smaller as $\Omega$ becomes of order 1 in the
radiation dominated close-to-flat-FL era.

We conclude that the numerics show  the occurance of residuals (i.e., 
definite spatial dependence is illustrated). We are not too concerned at this time
about the shape/characteristics of the residuals. However,
we can ask whether the numerics suggest void formation (see Fig. \ref{fig_waterfall}).
$\Omega$ develops a void at a spike location; they are voids
when they form. But it is hard to say what these imprints will lead to,
since they may change to overdensities via subsequent later dynamics.

%%\newpage

{\bf{Discussion}}.
We are interested in the possible existence of a GR
mechanism for generating matter perturbations.
We have concentrated on spikes, both incomplete spikes and
recurring spikes, and shown that there are  effects, entirely within classical GR, that could cause
inhomogeneities and hence matter perturbations.

In particular, we
have shown that there will be residual matter perturbations from spikes, based on
a heuristic qualitative analysis of a single spike in the OT $G_2$ model and
exact LRS $G_2$ and LTB spike solutions with matter \cite{thesis}, and most importantly from
numerical simulations. There are residuals from an incomplete spike,  
 that might in principle be large and thus affect structure formation (and any such effects 
might lead to observational constraints).

In addition, we have explicitly shown that there exist $G_2$ recurring spikes that lead to
inhomogeneities and a residual in the form of matter perturbations, that these occur
naturally within generic cosmology models within GR, and that they are not local but form
on surfaces and give rise to a distribution of perturbations.  In the $G_2$ models the
inhomogeneities can occur on a surface, and in general spacetimes the inhomogeneities can
occur along a line, leading to matter inhomogeneities forming on walls or surfaces.
Indeed, there are tantalising hints that (from dynamical and numerical analyses) that
filamentary structures and voids would occur naturally in this scenario.

Also there are shock waves that could generically
form (generated by inhomogeneities in the fluid pressure and density,
even within inhomogeneous Newtonian cosmology), not necessarily associated with spikes,
and leave a residual imprint in the matter perturbations. Incomplete spikes, or shock waves
not associated with spikes, might leave isolated (and perhaps large) perturbations.
However, shocks might form associated with spikes (indeed, spikes themselves will likely
cause shock waves which will, in turn, give rise to additional inhomogeneities 
and hence matter perturbations), and there might be interesting
behaviour due to interaction of a spike and a shock wave forming close to a spike,
leaving additional residual perturbations \cite{thesis}.

Inflationary cosmology provides a causal mechanism which 
generates the primordial perturbations which were later responsible for the formation of stars, galaxies, clusters, and 
all large scale structures of our Universe under the influence of gravitational collapse.
The density perturbations produced during inflation are due to
quantum fluctuations in the matter and gravitational fields \cite{infl}. 
The evolution
outside the Hubble radius then produces a large amplification of the perturbations. 
Primeval fluctuations are then thought to be present at the end of the inflationary
epoch.
Provided that
inflation lasts sufficiently long, 
generically an (almost) 
scale-invariant Harrison-Zel'dovich spectrum
of density fluctuations are generated, which then evolve to the
(tiny, $10^{-5}$) adiabatic, Gaussian, and scale invariant
density fluctuations in the power spectrum of the CMB \cite{infl}
(which then allows for a detailed comparison with current
observations).

If inflation does occur after the mixmaster regime is over, all of the classical
inhomogeneities will be redshifted away, but any `intermediate' scale effects might
not necessarily be redshifted away, and
might still be of importance in cosmology.
In the standard scenario it is assumed that there is no scale between
the quantum scale and the classical scale, that would be amplified during inflation
and become important in cosmology after inflation. We could ask whether there is such an 
intermediate  scale inherent in GR.

Perhaps the scale associated with distributed recurring spikes might be such an
intermediate scale. Indeed, since the `horizon' goes to zero as we approach the initial spacelike
mixmaster `singularity' in the classical regime, the scale (wavelength) of the `first'
recurring spikes can be arbitrarily small (but larger than the Planck scale assuming
the mixmaster oscillations occur in the classical regime after leaving the Planck
regime).
If there are classical scale inhomogeneities 
produced (either by late time, isolated spikes or shocks, or amplified recurring spikes)
they must be consistent
with current  CMB observations.

Indeed, in the case of recurring spikes within GR, the perturbations can be extremely
small (and, as noted earlier, occur in non-isolated distributions everywhere).
Therefore,  we
speculate whether recurring spikes could 
generate  primordial matter perturbations
that are very small  ($10^{-5}$)  at recombination 
and subsequently seed the large-scale structure of the actual universe, and consequently
act as an alternative
to the usual inflationary mechanism. 
In spite of the remarkable success of the inflationary Universe paradigm, there are several serious 
conceptual problems for current models  \cite{Brandenberger}. Therefore, a classical
GR mechanism for generating small inhomogeneities would be an interesting alternative,  
and might perhaps bring GR back to the centre of  cosmology \cite{Peebles}.
This is, of course, highly
speculative, but it might motivate the further study of spikes in GR.

Therefore, in future work we shall study what happens in more generality, hopefully obtaining
 some information
regarding the properties of residual matter inhomogeneities (e.g., 
their characteristic scales)  and the statistical properties
of matter perturbations (i.e., the 
distribution of perturbations in general inhomogeneous models).
In $G_2$ models perturbations occur on planar surfaces, while 
in general inhomogeneous ($G_0$) models perturbations may
occur on other types of surfaces.

Unfortunately,
although this is very interesting,
this is also extremely difficult to investigate, since it involves the
numerical integration of general GR cosmological models.  The numerical challenges include a
need to develop a numerical code capable of simulating spacetimes at super-horizon scales
while at the same time able to resolve spikes at sub-horizon scales, and to develop a
similar numerical code capable of handling shock waves.

{\em{Acknowledgment}}:
This work was supported, in part, by NSERC of Canada.

\end{document}